\documentclass[%
 reprint,
superscriptaddress,
 amsmath,amssymb,
 aps,
 prl,
]{revtex4-2}
\usepackage{xcolor}
\usepackage{graphicx}
\usepackage{dcolumn}
\usepackage{bm}

\begin{document}

\preprint{}
\title{Magnetic field generation in laser-solid interactions at strong-field QED relevant intensities}

\author{Brandon K. Russell}
 \email{bkruss@umich.edu}
\affiliation{%
G\'{e}rard Mourou Center for Ultrafast Optical Science, University of Michigan, 2200 Bonisteel Boulevard, Ann Arbor, Michigan 48109, USA
}

\author{Marija Vranic}%
\affiliation{%
 GoLP/Instituto de Plasmas e Fus$\tilde{a}$o Nuclear, Instituto Superior T\'{e}cnico, Universidade de Lisboa, 1049-001 Lisbon, Portugal
}%

\author{Paul T. Campbell}
\affiliation{%
G\'{e}rard Mourou Center for Ultrafast Optical Science, University of Michigan, 2200 Bonisteel Boulevard, Ann Arbor, Michigan 48109, USA
}

\author{Alexander~G.~R.~Thomas}
\affiliation{%
G\'{e}rard Mourou Center for Ultrafast Optical Science, University of Michigan, 2200 Bonisteel Boulevard, Ann Arbor, Michigan 48109, USA
}

\author{Kevin M. Schoeffler}%
\affiliation{%
 GoLP/Instituto de Plasmas e Fus$\tilde{a}$o Nuclear, Instituto Superior T\'{e}cnico, Universidade de Lisboa, 1049-001 Lisbon, Portugal
}%

\author{Dmitri A. Uzdensky}
\affiliation{
Center for Integrated Plasma Studies, Physics Department, 390 UCB, University of Colorado,
Boulder, CO 80309, USA
}

\author{Louise Willingale}
\affiliation{%
G\'{e}rard Mourou Center for Ultrafast Optical Science, University of Michigan, 2200 Bonisteel Boulevard, Ann Arbor, Michigan 48109, USA
}

\date{\today}

\begin{abstract}
 Magnetic field generation in ultra-intense laser-solid interactions is studied over a range of laser intensities relevant to next-generation laser facilities ($a_0 = 50-500$) using 2D particle-in-cell simulations. It is found that fields $\mathcal{O}$(0.1 MT) (1 GigaGauss) may be generated by relativistic electrons traveling along the surface of the target.
 However a significant fraction of the energy budget is converted to high-energy photons, $\sim38\%$ at $a_0=500$, greatly reducing the available energy for field generation. A model for the evolution of the target-surface fields and their scaling with $a_0$ is developed using laser parameters and assumed values for the average radial electron velocity and reflectivity.
The model and empirical scaling allow for the estimation of field strengths on the next generation of laser facilities, a necessary component to the proposal of any future magnetized experiment. 
\end{abstract}

\maketitle

Magnetic fields are integral to many high-energy-density physics (HEDP) experiments, e.g., the study of magnetized shocks \cite{Schaeffer_PRL_2017}, magnetic reconnection \cite{Nilson_PRL_2006,Raymond_PRE_2018,Palmer_POP_2019}, and magnetized turbulence \cite{Tzeferacos_NatCom_2018}.
These studies, often using high-power lasers, are generally motivated by astrophysical systems where magnetic fields are predicted to grow via Biermann battery \cite{Biermann_PR_1951}, magnetic dynamos \cite{Brandenburg_PR_2005}, or instabilities \cite{Schlickeiser_ApJ_2003}. In the laboratory, plasma motion may self-generate magnetic fields, or conducting coils can impose magnetic fields external to the plasma \cite{Diado_PRL_1986}.

For a laser pulse focused onto a solid target, strong magnetic fields can be generated. The strength and spatial profile of the fields varies greatly depending on the target material, density profile, and laser parameters.
Various experiments have been performed at moderate laser intensities ($\sim10^{14}-10^{15}$ Wcm$^{-2}$) using long-pulse (ns) lasers and have generated fields on the order of 100~T \cite{Nilson_PRL_2006,Gao_PRL_2015}.
In this non-relativistic regime, the field generation is well characterized by the Biermann battery mechanism ($\partial B/\partial t \propto \nabla T_e \times \nabla n_e$) \cite{Biermann_PR_1951,Li_PRL_2006,Schoeffler_PRL_2014,Campbell_PRL_2020}. 

At higher intensities, where the laser electric field becomes large enough to accelerate electrons to relativistic energies in excess of their rest mass, i.e.\ the normalized field strength $a_0 = eE/m_e\omega_L c \approx \sqrt{I\lambda_{\mu m}^2/1.4\times10^{18}(\textrm{Wcm}^{-2}\mu\textrm{m}^2)} >1$, magnetic field generation is not well understood. Solid target experiments have been performed in this regime, measuring field strengths $\mathcal{O}$ $(10\; \rm kT)$ using particle deflectometry \cite{Schumaker_PRL_2013,Sarri_PRL_2012,Palmer_POP_2019} and polarimetry of self-generated laser harmonics \cite{tatarakis_NAT_2002}.
Shukla \textit{et al.} performed particle-in-cell (PIC) simulations with $a_0\approx 1$ and found that Biermann magnetic field generation was dominant in the expanding plasma~\cite{Shukla_PRR_2020}.
Other recent studies of field generation schemes include the amplification of seed magnetic fields \cite{Shi_2020}, and the formation of strong magnetic fields along a channel wall by an intense $>10^{22}$ Wcm$^{-2}$ pulse interacting with a structured target \cite{Wang_PoP_2019}. A method for measuring fields for intensities exceeding $10^{23}$ Wcm$^{-2}$ interacting with near-critical density targets using the spin of ejected electrons has been proposed \cite{Gong_PRL_2021}.  

As the next generation of laser facilities pushes to even higher laser intensities, perhaps $>10^{23}$~Wcm$^{-2}$ \cite{Yoon_2021}, field generation will become more complex as quantum electrodynamic (QED) effects, i.e.,\ nonlinear Compton emission and pair creation, could influence the dynamics of the system \cite{Bell_PRL_2008,Ridgers_PRL_2012,Zhang_NJP_2015,Zhang_PoP_2020}.
Understanding how magnetic fields form and scale with laser intensity in this ultra-intense regime will be fundamental to any magnetized experiment performed using solid targets at these laser facilities. 

Here, we study magnetic field generation in the ultra-high intensity regime using QED PIC simulations over a broad range of laser intensities ($a_0 = 50-500$).
A scaling for the surface fields formed outside the focal volume as a function of $a_0$ is found and equations are derived to model the interaction allowing for the estimation of field strengths from experimental parameters. These surface fields outside of the focal volume are the fields that would interact in a standard two-beam magnetic reconnection geometry and are therefore the most relevant to future magnetized plasma experiments \cite{Nilson_PRL_2006}.

\begin{figure}\label{Fig:propagation}
    \includegraphics[]{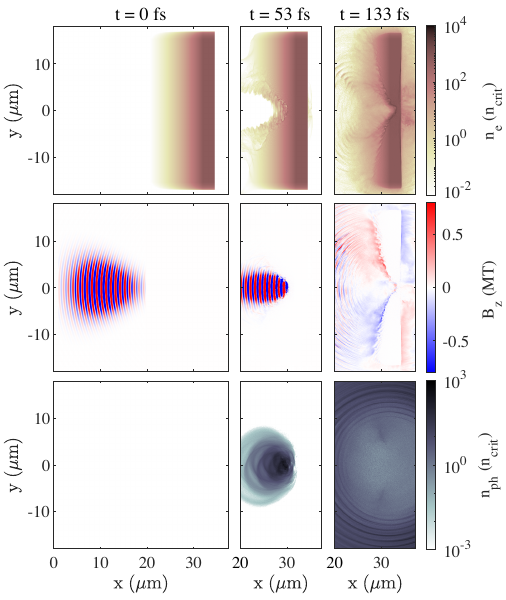}
\caption{Evolution of electron density $n_e$, out-of-plane magnetic field $B_z$, and photon density $n_{ph}$ for $\gamma$-ray photons with energies $>2m_ec^2$ from the interaction of a 20 fs, $a_0=500$ laser pulse with solid density target consisting of electrons and Al$^{13+}$ in a 2D OSIRIS simulation.}
\label{fig:fig1}
\end{figure}

Simulations were performed using the PIC code OSIRIS, which includes a Monte-Carlo module for the nonlinear Compton and Breit-Wheeler strong-field QED processes in the locally constant-field approximation (LCFA) \cite{Grismayer_PoP_2016,Grismayer_PRE_2017}. A 2D rectangular box with dimensions $37.4~\mu \rm{m} \times 36~\mu$m was used. A 20 fs full-width-half-max (FWHM), $1~\mu m$ central wavelength, laser pulse linearly polarized in the $y$-direction was initialized in the simulation box and propagated in the +$x$-direction. The pulse was focused with a FWHM of $3\;\mu$m onto a target consisting of electrons and Al$^{13+}$ at $x \approx 31.8 $ $\mu$m. Open boundary conditions were used for particles and fields on all boundaries. The particles were initialized with a peak charge density of a fully ionized solid target ($700n_{crit}$ or $\sim 7.8\times 10^{23}$~cm$^{-3}$) over a thickness of 2.5~$\mu$m. An exponentially decaying density ramp was initialized as the front edge of the target with a characteristic decay length of 1~$\mu$m to approximate a pre-plasma. While pre-plasma formation at these extreme intensities has not be characterized experimentally, simulations without any preformed plasma would not represent realistic pre-pulse contrast conditions. The target was tapered to the boundary in the $y$-direction to minimize boundary effects. The box was initialized with resolutions $\Delta x = 10.6$~nm and $\Delta y = 22.7$~nm (slightly larger than the collisionless skin-depth $c/\omega_p \approx 6$ nm at 700$n_{crit}$), generating a grid with $3525\times 1582$ cells. Each particle species was initialized with 400 particles-per-cell. Simulations were performed with a range of $a_0$ from 50 to 500 corresponding approximately to intensities of $3.5\times 10^{21}$~Wcm$^{-2}$ to $3.5\times 10^{23}$~Wcm$^{-2}$. This range of $a_0$ is well within the range of validity for the LCFA. Although the low energy part of the emitted photon spectra can be overestimated by the LCFA, in these simulations radiation reaction was set to be classical for particles with $\gamma < 10$, therefore these results are valid. To provide the temporal resolution required by the QED module, the timestep was set to 1/500$\omega_L^{-1}$ for $a_0 = 400$ and 500, and 1/267$\omega_L^{-1}$ for $a_0\leq267$, where the laser frequency $\omega_L = 1.885\times 10^{15}$ rad/s. These values are smaller than the time step required for the CFL condition (0.06$\omega_L^{-1}$) \cite{CFL}.

\begin{figure}
    \includegraphics[]{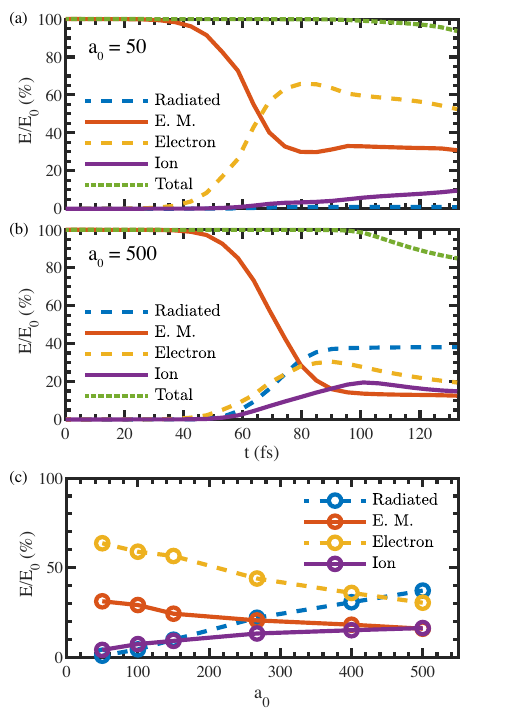}
\caption{Energy budget of 2D simulations at (a) $a_0 = 50 $ and (b) 500, and (c) the scaling of each component of energy with $a_0$ at $t = 90.2$ fs. The energy $E$ is normalized to the initial energy in the simulation box $E_0$. Only energy within the simulation box is included, therefore $E_{total}<100\%$ once energy leaves the box.}
\label{fig:budget}
\end{figure}

Fig.\ \ref{fig:fig1} shows the dynamics of the $a_0=500$ simulation through 2D maps of plasma density, magnetic field, and emitted photon density at 3 time snapshots. As the laser pulse interacts with the target, it rapidly accelerates the electrons from the central part of the target in all directions. The hot electrons generate strong out-of-plane magnetic fields $B_z > 0.1 $ MT on the front and back surfaces of the target. The spatial profile of the front-surface magnetic fields is not smooth. This is due to a combination of modulation of the electron density by the laser and the development of shear or streaming instabilities due to the large return current that forms. 

Once the laser pulse reaches the relativistic critical density ($n_{crit} = \langle \gamma \rangle m_e \varepsilon_0 \omega_L^2 / e^2$), a fraction of the pulse is reflected. This establishes a standing wave that interacts with hot electrons in the focal volume to generate energetic radiation. The photons shown (Fig.\ \ref{fig:fig1} row 3) have energies $>2m_ec^2$ (see Supplemental Fig.\ \ref{fig:spectra} for electron and photon spectra). While radiation is emitted in all directions, there is a bias towards the reflected laser direction. Since highly-relativistic electrons emit photons along their direction of motion, this radiation pattern reflects the underlying angular distribution of the accelerated electrons.

Fig.\ \ref{fig:budget} shows the time evolution of the energy budget for $a_0 = 50$ and 500 and the scaling of overall energy partition with $a_0$. The laser pulse is initialized in the simulation domain and defines the initial total energy of the system. As it interacts with the target, the laser energy is initially converted into electron kinetic energy, which is subsequently converted to ion kinetic energy, radiated energy (high energy photons), and the electric and magnetic fields on the target. The final electromagnetic field energy (E.~M.\ in the plot legend) is made up of the contributions from all electromagnetic fields in the simulation box. To decouple the laser field from the target fields, the fields outside the target surface were integrated at a time after the interaction was over, but before the reflected laser exited the box. This gave a reflectivity of $\sim22\%$ at $a_0 = 100$, and $\sim14\%$ at $a_0 = 400$, showing a weak dependence on $a_0$. The $E/E_0$ trends in Fig. \ref{fig:budget}(c) are taken at $t = 90.2$~fs, a time when the majority of energy still exists in the simulation box, i.e. $ E_{total}/E_0\approx100 \%$ [Fig.\ \ref{fig:budget}(a-b)]. These trends show an increase in radiated energy with increasing $a_0$. Conversely, the electron kinetic energy decreases with increasing $a_0$ due to radiative cooling and makes up a smaller fraction of the total energy than radiation at $t = 90.2$ fs for $a_0 = 500$. Pair creation makes up a negligible fraction of the energy budget with only $0.054\%$ at $a_0 = 500$. 

\begin{figure}
    \includegraphics[]{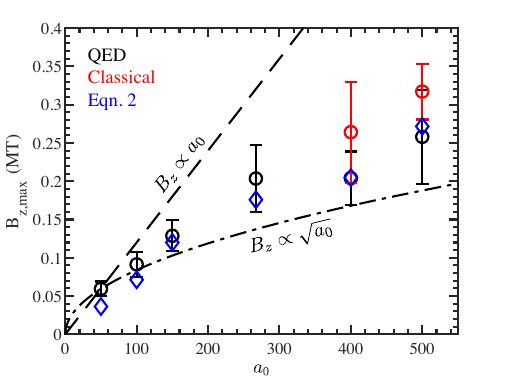}
\caption{Scaling of the maximum magnetic field $B_{z,max}$ as a function of $a_0$. The black circles are calculated from simulations including QED effects, taking an average of maximum $B_z$ along the target surface from $y = 8 - 12 \; \mu \rm{m}$. Error bars are the standard deviation of $B_{z,max}$. Red circles are the mean values in simulations without radiation effects. Eqn.\ \ref{eqn:Bz} was used with values of $x_{B,max}$ and $\langle u_y \rangle$ extracted from the simulations (blue diamonds).} 
\label{fig:scaling}
\end{figure}

\begin{figure}
    \includegraphics[]{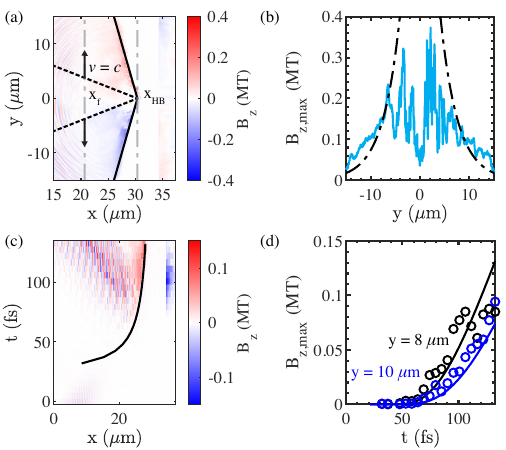}
\caption{An example of the simple model being used to calculate the surface magnetic fields for $a_0 = 100$. (a) Fit to surface where $B_z$ is maximum from Eqn.\ \ref{eqn:linsurf}, where $x_{HB}$ is calculated from Eqn.\ \ref{eqn:HB} using R = 0.22. (b) Blue line shows $B_{z,max}$ along $y$ at $t = 133$ fs, dash-dot black line shows the result of Eqn.\ \ref{eqn:Bz} from the surface in (a). (c) Space-time plot of $B_z$ at $y = 8 \mu$m and the result of Eqn.\ \ref{eqn:linsurf} (black line). (d) Evolution of $B_{z,max}$ (black circles) within $\pm 1\mu$m  of the line in (c) and the result of Eqn.\ \ref{eqn:Bz} (black line) calculated along this line with $\langle u_y\rangle = 0.12c$. An additional blue line is shown for a lineout at $y = 10\;\mu$m.} 
\label{fig:examplecalc}
\end{figure}

To understand how surface magnetic field generation ($B_z$) scales with $a_0$, we extracted the maximum values of $B_z$ along the $x$-direction and took the mean of these values in the $y$-direction along the target surface between $y$ = 8 to 12 $\mu \rm{m}$ (black circles in Fig.\ \ref{fig:scaling}). These fields outside the focal volume have historically been used for characterizing the initial configuration for magnetic reconnection experiments in the standard two-pulse geometry and are therefore the most relevant for future experiments. Error bars show the standard deviation in this region. The position of the surface where the maximum $B$-fields are generated is pushed to the positive $x$-direction as $a_0$ increases. For reference, lines of $B_z\propto a_0$ and $B_z \propto \sqrt{a_0}$ are shown. The scaling of $B_z$ appears to reside between these two reference lines; however, it is closer to $B_z \propto \sqrt{a_0}$. Even with QED effects, for the range of simulated $a_0$, $B_z$ continues to grow with $a_0$, reaching maximum values $>0.25$ MT or 2.5 GG. If $a_0$ were to be increased further, it is unclear how the the scaling would change as pair production may become important. 

The effect of radiation reaction on field generation can be elucidated by running the simulations where photons make up a significant fraction of the energy budget, i.e. $a_0 = 400$ and 500, with QED effects and radiation reaction switched off. The results are plotted as red circles in Fig.\ \ref{fig:scaling}, which show a $\sim29\%$ and $\sim23\%$ increase in $B_{z,max}$ compared to the QED simulations for $a_0 = 400$ and 500 respectively. The dynamics of the interaction are quite similar outside the focal volume, however an interesting phenomenon is seen near the focal spot (see Supplemental Fig.~\ref{fig:Weibel}). Weibel filaments grow near the focal spot in the classical simulation, a process previously noted by Shukla \textit{et al.} for $a_0\approx 1$ \cite{Shukla_PRR_2020}, whereas in the QED simulation these filaments are suppressed. Understanding the effects of radiation reaction and QED on the Weibel instability will be the subject of a future study. 

To model the evolution of the fields and their scaling with $a_0$ analytically using parameters that can be measured experimentally, we develop the following simple theoretical model starting with the following equation obtained by Schumaker \textit{et al.} \cite{Schumaker_PRL_2013}:
\begin{equation}\label{eqn:schu}
B_z(x,y,t) \approx -\mu_0\int_{-\infty}^x j_y(x',y,t)dx'.
\end{equation}
This equation is derived from a combination of the Amp\`{e}re-Maxwell equation, Gauss's law, and the continuity equation assuming that that the displacement current normal to the target is the only non-negligible term in the Amp\`{e}re-Maxwell equation. Numerically integrating $j_y$ from the -$x$ boundary to a position $x$ was found to give a good approximation of $B_z$ at that position. The current cannot be measured in experiment, therefore it must be replaced with quantities that could be measured or estimated. As the laser interacts with the target, it channels and hole-bores, accelerating electrons outward, primarily along the surface of the target, inducing a return current. This causes the target to contract, leaving behind the quasi-stationary ions. The outward flowing electrons respond to the target-normal electric fields, forming a density profile similar to the initial ion density profile. The current can then be approximated using a spatially averaged electron velocity $\langle u_y \rangle = \langle j_y/en_e \rangle$ multiplied by the initial ion charge density profile $\rho_{i0} = en_0\exp{[(x - x_0)/L]}$. Substituting it into Eqn.\ \ref{eqn:schu} and integrating this current to the position $x_{B,max}$ of the magnetic field maximum gives:
\begin{multline}\label{eqn:Bz}
    B_{z,max}(y,t) \approx \\ e\mu_0\langle u_y\rangle n_0L\exp\left(\frac{x_{B,max}(y,t) - x_0}{L}\right ),
\end{multline}
where $L$ is the characteristic decay length of the pre-plasma and $x_0$ marks the transition point from pre-plasma to bulk density.

The position $x_{B,max}$ can be approximated by noting that the laser drives the electrons outward at $\sim c$ and an approximately linear surface is formed as the channel expands. This surface is defined by two points, one at a position $x_f$ that propagates outward in $y$ at $c$, and the other fixed at $x_{HB}$, the position the laser hole bores to [see Fig.~\ref{fig:examplecalc}(a) for diagram]:
\begin{equation}\label{eqn:linsurf}
    x_{B,max}(y,t) = x_f - \frac{[|y| - c(t - t_0)](x_{HB} - x_f)}{c(t - t_0)},
\end{equation}
\begin{multline}\label{eqn:HB}
     x_{HB} = x_0 + L\ln(\frac{\gamma_{osc}n_{crit}}{n_0}) \\ + 2L\ln\left[1 + \frac{c}{L}\sqrt{\frac{\ln(2)\tau^2\gamma_{osc}}{4R_{qm}} }    \right], 
\end{multline}
 where $R_{qm} = M_i/Zm_e$, $\gamma_{osc} = \sqrt{1 + (1+R)a_0^2/2} \gg 1$, $R$ is the reflectivity and $x_f$ and $t_0$ are fitting parameters. Eqn.\ \ref{eqn:HB} describes the hole boring position for a laser with the envelope $a(t)\propto \exp(-2t^2/\tau^2)$. This is modified from the equation of Kemp \textit{et al.} \cite{Kemp_PRL_2008} assuming the laser hole bores into an exponential density profile for the FWHM duration of the pulse with a constant normalized vector potential $a = a_0$. Eqn.\ \ref{eqn:linsurf} is a linear fit to the maximum-$B_z$ surface.
 
 An example calculation for $a_0 = 100$ is shown in Fig.~\ref{fig:examplecalc}. The position $x_{B,max}$ is accurately fit through Eqn.~\ref{eqn:linsurf} where $x_f$ is taken to be the position where the exponential preplasma is truncated, and $t_0$ was varied until a rough fit was achieved. The value used for $t_0$ (20~fs) is slightly delayed from the time at which the laser begins to interact with the pre-plasma ($\sim$5 fs). These same fitting parameter values could be used with good accuracy for all tested $a_0$. The maximum $B_z$ along this line was calculated from Eqn.\ \ref{eqn:Bz} and compared to the simulated values in Fig.~\ref{fig:examplecalc}(b). The average electron $y$ velocity $\langle u_y \rangle$ was calculated by taking an average of $j_y/en_e$ over the area defined by the positions in the $y$-direction from $y = 8$ to $12 \mu$m and in the $x$-direction from $x_{B,max}-1$ $\mu$m to $x_{B,max}$, i.e. a 1 $\mu$m thick band from the surface of maximum $B_z$ (Supplemental Fig. \ref{fig:avg_uy} shows that $\langle u_y\rangle$ scales weakly with $a_0$). This average velocity is significantly less than $c$, contrary to what has been assumed in previous theory. The average velocity of electrons decreases significantly in regions of strong magnetization as their trajectories become affected by the magnetic fields. Similar $B_{z,max}$ magnitudes are seen outside of the focal volume; however, inside the focal volume the fields are overestimated. The equations become invalid in this region as ion motion becomes important and the surface is curved and cannot be approximated by a linear fit. However, outside the focal volume the equations provide an accurate estimate for the position [Fig.\ \ref{fig:examplecalc}(c)] and magnitude [Fig.~\ref{fig:examplecalc}(d)] of the evolving fields. Additionally, using Eqn.~\ref{eqn:Bz} with simulated values for $x_{B,max}$ and $\langle u_y\rangle$ a similar scaling with $a_0$ is found (Fig. \ref{fig:scaling} blue diamonds). 
 
 Additional simulations were performed with varying scale lengths ($L$ = [1~$\mu$m, 0.5~$\mu$m, 0.25~$\mu$m] at $a_0 = 100$). For these simulations the early evolution of the fields can be accurately calculated using the same $t_0$ and changing the position $x_f$ to the position where each simulated preplasma is truncated. However the field generation in the short scale-length simulations is similar to the rear-surface field generation where the strongest magnetic fields rapidly propagate out radially (see Supplemental Fig. \ref{fig:examplecalc2}). 
 
Taken together, Eqns. \ref{eqn:Bz}, \ref{eqn:linsurf}, and \ref{eqn:HB} provide a way to estimate the maximum $B_z$ at some radial position $y$ on the target surface at some time $t$ during the expansion. This estimate can be made directly from $a_0$ and laser pulse duration $\tau$, therefore this equation set provides significant utility in estimating field strengths for future experiments on the next generation of lasers. Aside from the laser parameters, values for $\langle u_y\rangle$ are assumed, which can be taken to be in the range of 0.2c to 0.1c for increasing $a_0$ from 50 to 500. Additionally, $R$ may be measured experimentally or calculated from the equations presented by Zhang \textit{et al.}\ \cite{Zhang_NJP_2015}.  
 
With the $>0.1$ MT fields produced outside the focal volume, it may be possible to perform experiments studying magnetized plasma processes, e.g. reconnection, shock formation, turbulence, in regimes relevant to the extreme plasma environments around relativistic compact objects such as neutron stars and black holes. For example, two ultra-intense lasers may be used in the standard 2-beam geometry \cite{Nilson_PRL_2006} to study highly energetic laser-driven magnetic reconnection. The regime accessed by such an experiment can be parameterized using the magnetization parameter $\sigma = B^2/\mu_0w$ which is the ratio of magnetic enthalpy density to relativistic plasma enthalpy density $w$. The latter is the sum of the plasma pressure and internal energy (including the rest-mass energy density). For $a_0 = 400$, the electron-only magnetization $\sigma_e$, calculated in the co-moving frame of hot electron outflow from the focal volume, is in the range of 0.1-0.5 along the surface of the target. This is quite large, allowing for semi-relativistic outflows from the reconnection region and giving the already ultra-relativistically hot electron inflow a modest amount of additional energy. With the production of such energetic conditions, next-generation laser facilities will be promising platforms for the study of the magnetized processes that occur in extreme astrophysical environments.

This work is supported by the National Science Foundation through award number 1751462. The authors would like to acknowledge the OSIRIS Consortium, consisting of UCLA and IST (Lisbon, Portugal) for providing access to the OSIRIS 4.0 framework which is supported by NSF ACI-1339893. D.A.U. was supported by NSF grants AST-1806084 and AST-1903335. A.G.R.T was supported by NSF grants 2206059 and 2108075.

\bibliography{Magbib}

\clearpage

\section{Supplemental Material}

\subsection{Radiation Spectra}

\begin{figure}[h]
    \includegraphics[]{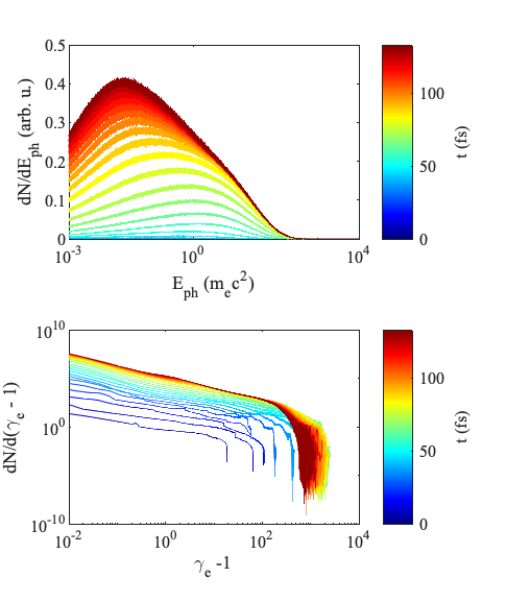}
\caption{Time evolution of photon and electron energy spectra integrated over the simulation box for $a_0 = 500$. Here, $E_{ph}$ is the photon energy normalized to the electron rest mass energy $(m_ec^2)$.}
\label{fig:spectra}
\end{figure}

 The photon and electron spectra from the $a_0 = 500$ simulation are shown in Fig.\ \ref{fig:spectra}. At this laser intensity, a significant fraction of photons have energies $>m_ec^2$ and electrons are highly relativistic. 
 
    \subsection{Suppression of Weibel Instability}

\begin{figure} [h]
    \includegraphics[]{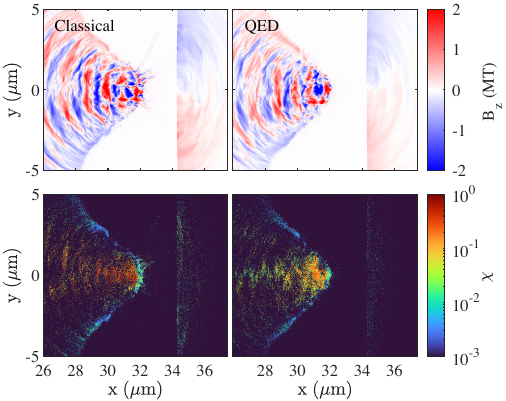}
\caption{Magnetic field and $\chi$ at t = 85~fs in simulations of $a_0 = 400$ with (QED) and without (classical) radiation reaction showing the suppression of filaments formed near the focal spot.} 
\label{fig:Weibel}
\end{figure}

  When observing the differences in simulations without radiation reaction (classical) and with radiation reaction (QED), we find that filaments formed near the focal spot are suppressed in the QED case. This is seen in Fig.\ \ref{fig:Weibel} in the magnetic field $B_z$. Notably, $\chi = |F^{\mu\nu}p_\nu|/m_eE_{cr}$, the ratio of the field strength in the frame of the electrons to the critical field $E_{cr} = m_e^2c^3/e\hbar = 1.3\times 10^{18}$ V/m is large within the filaments. Therefore, if filaments were to form, they may be suppressed by radiation reaction, or the electrons responsible for generating the filaments may lose their energy to radiation before entering the high-density part of the target. This phenomenon will be the subject of a future study. 
  
  \subsection*{Average $y$ velocity}
  
\begin{figure}[h]
    \includegraphics[]{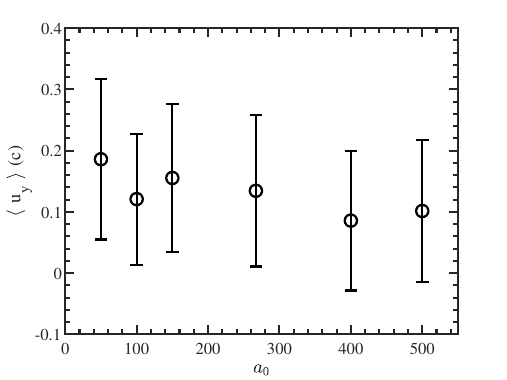}
\caption{Mean velocity in the $y$-direction outside the focal volume along the surface of the target.}
\label{fig:avg_uy}
\end{figure}

  The average velocities in the $y$-direction for various $a_0$ are shown in Fig.\ \ref{fig:avg_uy}. These values are calculated by taking the mean of $j_y/en_e$ in the region $y = [8,12] \; \mu$m and $x = [x_{B,max} - 1\;\mu\rm{m},x_{B,max}]$. In other words, it is the average velocity outside the focal volume within a region $1\;\mu$m from the surface where the magnetic field is maximum. The error bars show the standard deviation of the $y$-velocity in this region, which is quite large. However, the mean value of the velocity is approximately constant with varying $a_0$, only decreasing from $\sim0.2c$ to 0.1$c$ over a range of $a_0$ from 50 to 500. 
  
  \subsection*{Varying Scalelength}
  
    \begin{figure}
    \includegraphics[]{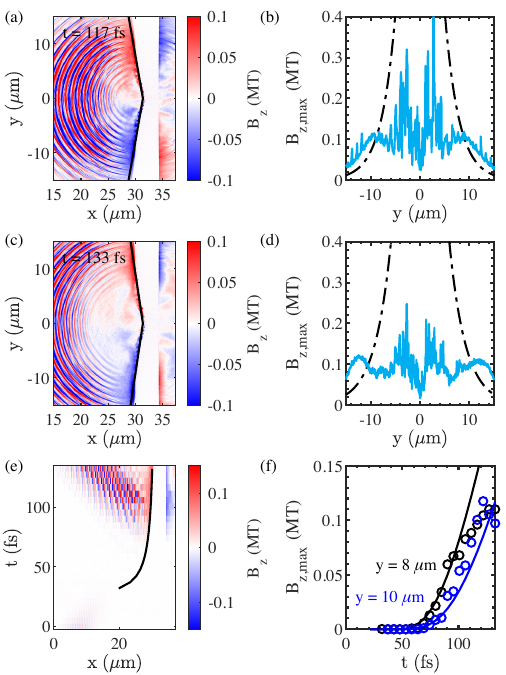}
\caption{The evolution of fields and comparison to the model presented in this paper for a  $0.5\;\mu$m scale-length. (a) and (c) show the fields with the surface $x_{B,max}$ fit from Eqn.\ \ref{eqn:linsurf} at $t = 117$~fs and $t = 133$~fs respectively. Plots (b) and (d) are the respective $B_{z,max}$ along the target surface with the black dash-dot line calculated from Eqn.\ \ref{eqn:Bz}. (e) shows the evolution of the fields and the surface fit at $y = 8\;\mu$m. (f) shows $B_{z,max}$ within $\pm1\;\mu$m of this line (black circles) and the calculated $B_{z,max}$ from Eqn.\ \ref{eqn:Bz} (black line). The evolution at $y = 10\;\mu$m is shown in blue.}  
\label{fig:examplecalc2}
\end{figure}

 An example of the simple model being used to calculate the surface magnetic fields for $a_0 = 100$ with a shorter $0.5\;\mu$m scale-length is shown in Fig.\ \ref{fig:examplecalc2}. A fit to the surface where $B_z$ is maximum from Eqn.\ \ref{eqn:linsurf} at $t = $ 117~fs is shown in (a). The blue line in (b) shows $B_{z,max}$ along $y$ at $t = 117$~fs and the dash-dot black line shows the result of Eqn.\ \ref{eqn:Bz} from the surface in (a). Similar to the calculation performed for the $1\;\mu$m scale-length, the model fits best outside the focal volume. Plots (c-d) are like (a-b) but taken at $t = $ 133~fs. At this time, the fields appear pulse-like, existing as a short burst of magnetic field that propagates along the surface of the target. As the magnetic fields begin exiting the box, $B_{z,max}$ no longer matches the theory. As the scale-length becomes smaller, the field generation becomes more similar to the rear-surface where there is no initial pre-plasma. Plot (e) shows the evolution of $B_z$ at $y = 8\; \mu$m and the result of Eqn.\ \ref{eqn:linsurf} (black line). The black line closely follows the surface evolution, therefore the surface is still well approximately by a linear fit at this smaller scale-length. The evolution of $B_{z,max}$ within $\pm 1 \; \mu$m  of the line in (c) is plotted as black circles in (f) and the result of Eqn.\ \ref{eqn:Bz} calculated along this line with $\langle u_y\rangle = 0.26c$ is plotted as a black line. Values calculated from a lineout at $y = 10\;\mu$m are shows as blue circles and the calculated values are shown as a blue line. This again shows that the evolution is well approximated by the equations, however at late times the theory begins to over-estimate the maximum field strength because the current has decayed and therefore the magnetic field is reduced. The average $y$-velocity is notably larger here, and therefore varies with scale-length.

\end{document}